\documentclass[%
10pt,
twocolumn,
amsmath,
amssymb,
amsfonts,
aps,
prb,
floatfix,
]{revtex4-2}
\usepackage[utf8]{inputenc}
\usepackage[T1]{fontenc}

\usepackage{bm}
\usepackage{mathrsfs}
\usepackage{xspace}

\usepackage{graphicx}
\usepackage{dcolumn}

\newcommand{\tens}[1]{\bm{\mathsf{#1}}}
\newcommand{\tenscomp}[1]{\mathsf{#1}}

\renewcommand{\i}{\mathrm{i}}
\newcommand{\e}{\mathrm{e}}
\renewcommand{\k}{\bm{k}}
\newcommand{\I}{\tens{I}}
\newcommand{\q}{\bm{q}}
\newcommand{\R}{\bm{R}}
\newcommand{\G}{\tens{G}}
\newcommand{\V}{\tens{V}}
\newcommand{\A}{\tenscomp{A}^{T_0}}
\newcommand{\N}{\tens{N}}
\newcommand{\Nt}{\tilde{\N}}

\newcommand{\Ncomp}{\tenscomp{N}}
\newcommand{\Ntcomp}{\tilde{\Ncomp}}
\newcommand{\Wcomp}{\tenscomp{\Omega}}

\newcommand{\Neq}[1]{\bar{\Ncomp}^{#1}}
\newcommand{\Nteq}[1]{\tilde{\bar{\N}}^{#1}}
\newcommand{\deltaN}{\delta\tilde{\bar{\N}}}
\newcommand{\deltaNcomp}{\delta\tilde{\bar{\Ncomp}}}

\newcommand{\W}{\tens{\Omega}}
\newcommand{\Lin}{\mathcal{L}}
\newcommand{\Lint}{\tilde{\Lin}}
\newcommand{\Gr}{\mathcal{G}}
\newcommand{\Grt}{\tilde{\Gr}}
\renewcommand{\P}{\tens{Q}}
\newcommand{\Pt}{\tilde{\P}}
\newcommand{\PtgradT}{\Pt_{\nabla T}}

\DeclareMathOperator{\Tr}{Tr}
\DeclareMathOperator{\vecop}{vec}

\newcommand{\CPB}{CsPbBr$_3$\xspace}
\newcommand{\LZO}{La$_2$Zr$_2$O$_7$\xspace}

\begin{document}
\title{Transition from population- to coherence-dominated nondiffusive thermal transport}
\author{Laurenz Kremeyer}
\email{laurenz.kremeyer@mail.mcgill.ca}
\affiliation{Department of Physics, Centre for the Physics of Materials, McGill University, Montreal, Canada}
\author{Bradley J. Siwick}
\affiliation{Department of Physics, Centre for the Physics of Materials, McGill University, Montreal, Canada}
\affiliation{Department of Chemistry, McGill University, Montreal, Canada}
\author{Samuel Huberman}
\email{samuel.huberman@mcgill.ca}
\affiliation{Department of Chemical Engineering, McGill University, Montreal, Canada}
\affiliation{Department of Physics, McGill University, Montreal, Canada}
\begin{abstract}
Deviations from diffusive heat transport in high thermal conductivity crystalline insulators are generally understood within the framework of the phonon Boltzmann Transport Equation.
However, for low thermal conductivity materials with large primitive cells or strong anharmonicity, the recently developed Wigner Transport Equation is more appropriate as it includes tunneling between overlapping phonon bands.
In this work, via solutions to the Wigner Transport Equation, we develop a scheme to obtain the dynamics of the phonon populations and coherences as a function of an arbitrary heat source.
The approach is applied to predict size effects and dynamical thermal conductivities in \CPB and \LZO using first-principles data as input.
We predict significant deviations from the bulk thermal conductivity in these materials at length scales on the order of hundreds of nanometers to a few microns at room temperature, well within the reach of direct observation using current experimental techniques.
\end{abstract}
\maketitle

\section{Introduction}
Diffusive heat transport in crystalline insulators occurs when the relevant transport space and/or time scales are much larger than the mean free paths (MFPs) and/or lifetimes of phonons.
As the system size approaches the MFPs, or the timescale associated with the heat source approaches phonon lifetimes, deviations from diffusive transport are expected.
Canonical examples are the reduction in thermal conductivity as the system size decreases~\cite{ju1999phonon,liu2006thermal,cheaito2012experimental,scott2021simultaneous} (sometimes referred to as the size effect) or as the size of the thermal excitation decreases~\cite{johnson2013direct,hu2015spectral,cuffe2015reconstructing,zeng2015measuring,huberman2017unifying}.
A more exotic, but unambiguously nondiffusive, behavior is phonon hydrodynamic transport, where momentum conserving phonon scattering processes dominate over resistive or boundary scattering and the spectacular phenomena of second sound can occur~\cite{huberman2019observation,jeong2021transient,ding2022observation,kremeyer2024ultrafast}.
In these nondiffusive regimes, the phonon Boltzmann transport equation (BTE) is typically used and has been shown to accurately model a range of transport behaviors.
While the BTE is suitable for high thermal conductivity materials, where the phonon linewidths are narrow and interactions are dominated by scattering mechanisms (i.e., phonon-phonon, boundary, defect etc.) with rates $\Gamma_s(\q)$ that are much less than the phonon frequency $\omega_s(\q)$, the BTE fails to capture the contribution of phonon wavelike tunneling between branches that can occur in low thermal conductivity solids.
tunneling-mediated inter-branch phonon transport becomes relevant when phonon linewidths are on the order of the inter-branch spacing (i.e., $(\Gamma_s(\q)+\Gamma_{s'}(\q))/2 \approx (\omega_{s'}(\q)-\omega_s(\q))$). This condition is generally satisfied in anharmonic materials with a large number of atoms in the unit cell, but also for glasses where the Brillouin zone reduces to the $\Gamma$ point and the same condition applies, $(\Gamma_s(\vec 0)+\Gamma_{s'}(\vec 0))/2 \approx (\omega_{s'}(\vec 0)-\omega_s(\vec 0))$.
When such tunneling is significant, the density matrix describing the phonon system cannot be assumed to be diagonal.
Phonon coherences, not just populations, become important and interference phenomena play an important role in determining transport in this regime.
It should be noted that in all cases discussed here, the vibrational modes do not fall below the Ioffe-Regel limit ($\Gamma_s(\q) = \omega_s(\q)$).
Recent work derived an expression for thermal conductivity which includes this coherence contribution based on the Wigner Transport Equation (WTE)~\cite{simoncelli2019unified}.
Here, we generalize these results to investigate the case of nondiffusive transport in the limit where inter-branch tunneling is significant.
We begin by proposing an extension to the WTE to encompass the effects of a space- and time-dependent heat source, which connects the framework to time-resolved experimental approaches that can be used to test the predictions of this theory. Following the same notation as~\cite{simoncelli2019unified}, the WTE is
\begin{widetext}
\begin{equation}
\frac{\partial}{\partial t} \N({\R},{\q},t)
+\i\Big[\W(\q),\N(\R,\q,t)\Big]+\frac{1}{2}\Big\{\V(\q), \vec{\nabla}_{\R} \N(\R,\q,t)\Big\}=\frac{\partial}{\partial t}\N(\R,{\q},t)   \bigg|_{\mathrm{H}^{\mathrm{col}} } + \P({\R},{\q},t), \label{eq:Wigner_evolution_equation_N}
\end{equation}
where $\N({\R},{\q},t)$ is the generalization of a one-phonon distribution (from a rank 1 to a rank 2 tensor) obtained via Wigner transformation of the time-evolution of the one-phonon density matrix.
$\R$ and $t$ are spatiotemporal coordinates and $\q$ is the phonon wave vector.
The collision term $\partial_t\N({\R},{\q},t) \big|_{\mathrm{H}^{\mathrm{col}} }$ describes phonon scattering and tunneling processes.
The diagonal elements, $\Ncomp_{s,s}({\R},{\q},t)$,  correspond to the conventional phonon populations of mode~$s$ described by the BTE and the off-diagonal elements correspond to phonon coherences between modes~$s$ and~$s'$ and $\Wcomp_{s,s'}(\q)=\omega_s(\q)\delta_{s,s'}$.
The tensor $\V(\q)$ is a generalization of a phonon's group velocity of rank 3 at every $\q$-point.
That leads to an implicit tensor contraction over Cartesian coordinates $\alpha$ in the anti-commutator expression
\begin{equation}
\frac{1}{2}\Big\{\V(\q), \vec{\nabla}_{\R} \N(\R,\q,t)\Big\} \equiv \frac{1}{2}\sum_\alpha \Big\{\V^\alpha(\q), \vec{\nabla}_{R^\alpha} \N(\R,\q,t)\Big\} = \frac{1}{2}\sum_\alpha \sum_{s''} \Big( V_{ss''}^\alpha \partial_{R^\alpha}N_{s''s'} + \partial_{R^\alpha} N_{ss''} V_{s''s'}^\alpha \Big)
\end{equation}
and analogously for the commutator.
As was done previously when solving the BTE~\cite{chiloyan2021green}, we have added the term $\P({\R},{\q},t)$ which describes the heat source.
For the purposes of generality, $\P({\R},{\q},t)$ is a matrix-valued function, where the diagonal terms correspond to the production of phonon populations and the off-diagonal terms allow for the production of phonon coherences.
While most thermal drives are well-described as purely diagonal excitations, the generation of coherent phonons, e.g., via impulsive Raman-type excitations, corresponds to off-diagonal source components of $\Pt$, given the excited modes share inter-mode tunneling channels~\cite{merlin1997generating,park2005mechanism}.

To solve Eq.~\ref{eq:Wigner_evolution_equation_N}, we take the following steps.
First, we Fourier transform the expression in space ($\R$) and time ($t$), denoted by $\mathscr{F}$, to obtain Eq.~\ref{eq:Wigner_evolution_equation_N}, where $\omega$ and $\k$ are the temporal and spatial Fourier variables:
\begin{equation}
-\i \omega \Nt({\k},{\q},\omega)+i\Big[\W(\q),\Nt({\k},{\q},\omega)\Big]+\frac{\i \k}{2}\Big\{\V(\q), \Nt({\k},{\q},\omega)\Big\} = \mathscr{F} \left\{\frac{\partial}{\partial t}\N({\R},{\q},t) \bigg|_{\mathrm{H}^{\mathrm{col}} }\right\} + \Pt({\k},{\q},\omega).\label{eq:Wigner_evolution_equation_N_FT}
\end{equation}
\end{widetext}
In the following, we drop the explicit dependence on $\R,\q,t$ or $\k,\q,\omega$, and instead indicate it by the use of a tilde above the respective symbol.
Following the operation of $\mathscr{F}$, spatial and temporal derivatives transform into simple algebraic terms and the WTE now takes the form $\Lint \Nt=\Pt$, where $\Lint$ is a linear operator combining all terms on the left-hand side and the collision operator on the right-hand side of Eq.~\ref{eq:Wigner_evolution_equation_N_FT}.
We will touch on the Fourier-transformed collision term later and discuss different assumptions and models that can be used to describe scattering in the system.
Regardless of the details of the chosen scattering operator, one can introduce a Green's function $\Grt(\k,\q,\omega)$ defined by
\begin{equation}
\Lint \Grt = \I\quad \longleftrightarrow \quad \Lin\Gr=\I\delta(\R)\delta(t), \quad\text{where}\quad \tenscomp{I}_{s,s'}=\delta_{s,s'}.
\end{equation}
The Green's function $\Grt$ encodes how the phonon system responds to an impulsive excitation that is localized in space and time.
Once $\Grt$ is obtained by inversion of $\Lint$, the solution for an arbitrary heat source $\Pt$ can be directly computed as $\Nt = \Grt \Pt$ for the pair of $(\k, \omega)$ for which $\Lint$ was inverted.
Computing the system's Green's function for a range of frequencies $\omega$ at a specific $\k$ provides a frequency-response picture of the system and allows us to recover the Wigner distribution at fixed wave vector $\k$ via the inverse Fourier transform in time
$\N(\k,\q,t) = (2\pi)^{-1}\int \mathrm{d}\omega\,\mathrm{e}^{-\i\omega t} \Nt(\k,\q,\omega).$
From the real-time Wigner distribution of the system, many experimentally accessible observables can be computed.
Pump-probe experiments are the first that come to mind, since they involve an ultrashort excitation of the studied system that very closely resembles the underlying Dirac distribution $\delta(t)$ of the Green's function approach.
Thus, the frequency-domain Green's function offers an intuitive, quantitative connection between this theory and time-resolved pump-probe data.

Explicitly calculating the inverse operator $\Lint^{-1}$ to obtain $\Grt$ can be computationally very expensive---especially for systems with complex unit cells.
It is faster to use solvers that do not explicitly compute the full Green's function while still solving a system of the form $\Lint \Nt=\Pt$, with the drawback that the system is only solved for a specific source term.
Accordingly, if responses are needed for many different sources, forming $\Grt=\Lint^{-1}$ can be more efficient overall; if only a few sources are of interest, solving $\Lint\Nt=\Pt$ per source is preferable.
The calculations presented here were performed using a mix of both approaches, as the code implementation supports easy switching between methods.

\section{Theory}
We first discuss the theory of calculating dynamical thermal conductivities and then move on to solving the WTE for arbitrary source terms.
Throughout this work, we invoke small deviations from equilibrium and linearize the local-equilibrium distribution around a reference temperature $T_0$, which is the system base-temperature before photoexcitation in a pump-probe experiment.
To first order, this means
\begin{equation}
\Nteq{T_l} = \Nteq{T_0}\delta(\k)\delta(\omega) + \deltaN,
\end{equation}
\begin{equation}
\Nt = \Nteq{T_0}\delta(\k)\delta(\omega) + \delta\Nt,
\end{equation}
with the diagonal matrix $\deltaN = \frac{\partial \Nteq{}}{\partial T}\big|_{T_0} \Delta T$ with its nonzero elements $\deltaNcomp_{s,s} = \frac{\mathcal{V}N_\mathrm{C}}{\hbar \omega_s}c_s \Delta T$.
The delta-distributions arise from the Fourier transform of the time- and space-independent equilibrium distribution.
All material parameters have been calculated only at the background temperature $T_0$.
It is assumed that those do not change significantly for $\Delta T \ll T_0$ and that $\delta\Nt \ll \Nteq{T_0}\delta(\k)\delta(\omega)$.

\subsection{Dynamical thermal conductivity}
For the collision term we choose a closed-form expression in the relaxation-time approximation (RTA)~\cite{callaway1959model, allen2013improved} that follows from the linearization about the equilibrium temperature $T_0$ and is expressed in terms of one-body quantities as
\begin{equation}
\begin{split}
\mathscr{F} \left\{\frac{\partial}{\partial t}\N(\q)\bigg|_{\mathrm{H}^{\mathrm{col}} }\right\}{=}-(1-\delta_{s,s'})\frac{\Gamma_{s}+\Gamma_{s'}}{2} \Ntcomp_{s,s'}(\q)\\
 -\frac{\delta_{s,s'}}{\mathcal{V}N_\mathrm{C}}{\sum\limits_{{s''}{\q''}}} \A_{s,s''}(\q,{\q''})\left(\Ntcomp_{{s''},{s''}}(\q'')-\mathscr{F} \left\{\Neq{T_l}_{s''}(\q'')\right\}\right), \label{eq:scattering_operator_FT}
\end{split}
\end{equation}
where $\mathcal{V}$ is the unit cell volume, $N_\mathrm{C}$ is the number of unit cells, $\mathscr{F} \left\{\Neq{T_l}_s({\q})\right\}{=}\mathscr{F} \left\{\left({\exp\left[\frac{\hbar\omega_s(\q)}{k_B T_l(\R,t)} \right]{-}1}\right)^{-1}\right\}$ is the Fourier transform of the Bose-Einstein distribution at local temperature $T_l(\R,t)$, $\Gamma_s(\q)=(\mathcal{V}N_\mathrm{C})^{-1}\A_{s,s''}(\q,\q)\delta_{s,s''}$ is the phonon linewidth and $\A_{s,s''}(\q,\q'')$ is the standard collision operator that accounts for anharmonicity and isotopic disorder~\cite{fugallo2013abinitio, cepellotti2016thermal}.
One can see that the collision operator does not couple populations to coherences.
Additionally, note that the collision term only annihilates coherences, while it both creates and annihilates populations.
Coherences are only created by the off-diagonal entries of the generalized velocity operator $\V$.
In the RTA, the collision term can be further simplified to
\begin{equation}
\mathscr{F} \left\{\frac{\partial}{\partial t}\N \bigg|_{\mathrm{H}^{\mathrm{col}} }\right\} = -\frac{1}{2}\left\{ \G, \Nt \right\} + \frac{1}{2} \left\{ \G, \Nteq{T_l} \right\},
\end{equation}
with $\G_{s,s'} = \delta_{s,s'} \Gamma_s$, where the first term damps the nonequilibrium Wigner distribution, while the second term ensures relaxation towards the \emph{local} equilibrium at temperature $T_l$.
Away from the zero-mode, for any $(|\k|,\omega) \in \mathbb{R}^2 \setminus \{(0,0)\}$, we can simplify to
\begin{equation}
\mathscr{F} \left\{\frac{\partial}{\partial t}\N \bigg|_{\mathrm{H}^{\mathrm{col}} }\right\} = -\frac{\Gamma_s+\Gamma_{s'}}{2}\, \Ntcomp_{s,s'}
  + \delta_{s,s'}\,\Gamma_s\deltaNcomp_{s,s},
\label{eq:scattering_operator_FT_RTA}
\end{equation}

To compute the dynamical thermal conductivity as a function of $\k$ and $\omega$, we study the linear response to a harmonic temperature field
\begin{equation}
T_l(\R,t) = T_0 + \Delta T \e^{\i((k\,\bm{e}_\beta )\cdot\R - \omega t)} \quad\text{with}\quad |\Delta T| \ll T_0.
\end{equation}
Greek indices $\alpha,\beta \in \{x,y,z\}$ denote Cartesian components of vectors and tensors.
We drive the system with a spatial harmonic along the direction $\beta$, i.e., $\k = k \bm{e}_\beta$.

We must choose a source term that (i) produces a response proportional to $\Delta T$, (ii) does not inject net power, and (iii) vanishes for a uniform $\Delta T$ (i.e., $\k=0$).
A convenient choice that meets all three requirements is
\begin{equation}
\PtgradT^\beta = \frac{\i k}{2} \{\V^\beta, \Nteq{T_0}\}\Delta T - \deltaN. \label{eq:gradT_source}
\end{equation}
The first term is the drift term of the local-equilibrium distribution under a gradient, i.e., the generalization of the Peierls-Boltzmann drift term, and the second term cancels the inhomogeneous part of the scattering operator in the RTA.
First, the chosen $\PtgradT$ is proportional to $\Delta T$ and by linearity of the WTE the response is also proportional to $\Delta T$.
Second, the second term of the source cancels the $\deltaN$ term in the RTA scattering operator; and since $\sum_{\q} \Tr \left[ \hbar \W \Pt^\beta \right]= 0$, no net power is injected.
Last, the net source, after cancellation with the RTA term, vanishes for $\k=0$ and we recover the expected equilibrium solution without drive.
Plugging Eqs.~\ref{eq:scattering_operator_FT_RTA} and \ref{eq:gradT_source} into Eq.~\ref{eq:Wigner_evolution_equation_N_FT} yields
\begin{equation}
\begin{split}
-\i \omega \Nt + \i \left[ \W, \Nt \right] + \frac{\i k}{2} \left\{ \V^\beta, \Nt \right\} + \frac{1}{2}\left\{ \G, \Nt \right\} \\= \frac{\i k}{2} \{\V^\beta, \Nteq{T_0}\} \Delta T
\end{split}
\end{equation}
From the resulting distribution $\Nt(\k,\omega)$ we can calculate the thermal conductivity from the heat flux $\bm{J}$ as
\begin{align}
\kappa^{\alpha\beta}(\k,\omega) &= \mathrm{Re}\left[ \frac{J^\alpha(\k,\omega)}{\i k^\beta \Delta T} \right] \\
 &= \mathrm{Re}\left[ \frac{\sum_{\q} \Tr\left[ \lbrace\W(\q),\V^\alpha(\q)\rbrace \Nt^\beta(\k,\omega,\q) \right]}{\frac{2\mathcal{V}N_\mathrm{C}}{\hbar}\,\i k^\beta \Delta T} \right].
\end{align}
Note that by choice of the gradient source term $\PtgradT$, we constructed a purely linear, inhomogeneous system with a known right-hand side.
The temperature field $T_l$ does not need to be determined self-consistently, as it is an input to the problem.
Without loss of generality, we can choose an arbitrary $\Delta T$ and directly obtain the thermal conductivity.

\begin{figure*}[t]
    \centering
    \includegraphics[width=\textwidth]{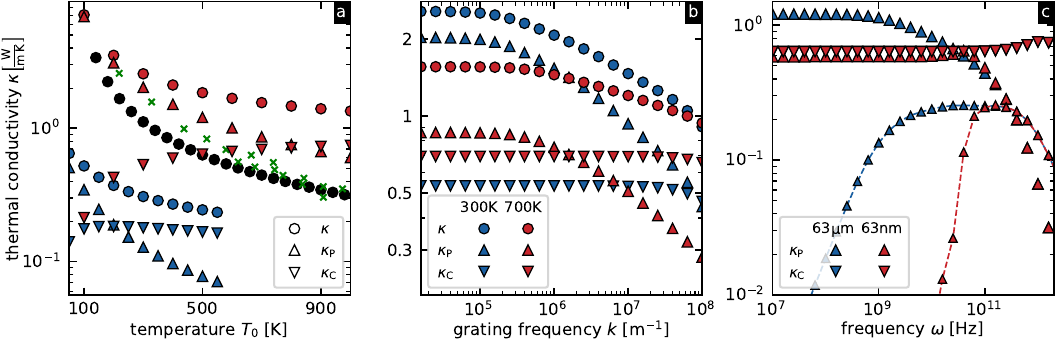}
    \caption{\textbf{a}\,Thermal conductivity of silicon (black), \CPB (blue), and \LZO (red) computed with the source term $\PtgradT$. The contribution to the overall thermal conductivity (round markers) from populations and coherences is shown with triangles pointing up and down, respectively. The thermal conductivity of silicon is scaled by a factor of 0.01 for visual clarity. Experimental reference data~\cite{glassbrenner1964thermal} is shown with green crosses. \textbf{b}\,Thermal conductivity of \LZO as a function of grating frequency at 300 and 700\,K, in blue and red respectively. \textbf{c}\,Dynamical thermal conductivity of \LZO at 500\,K for two different grating periods. Large markers denote the real part and smaller markers connected with a dashed line denote the imaginary part of the thermal conductivity. Note that $\mathrm{Re}[\kappa_\mathrm{C}]$ for both grating sizes fully overlap and that $\mathrm{Im}[\kappa_\mathrm{C}]$ are negligible.}
    \label{fig:reproduction}
\end{figure*}

\subsection{Arbitrary source terms}
The key difference in solving the WTE for arbitrary source terms is that net power can be injected into the system.
Plugging Eq.~\ref{eq:scattering_operator_FT_RTA} into Eq.~\ref{eq:Wigner_evolution_equation_N_FT} gives the WTE in the RTA for arbitrary source terms
\begin{equation}
\begin{split}
-\i \omega \Nt + \i \left[ \W, \Nt \right] + \frac{\i k}{2} \left\{ \V^\beta, \Nt \right\} + \frac{1}{2}\left\{ \G, \Nt \right\} \\
= \Pt + \frac{1}{2}\left\{ \G, \Nteq{T_l}\right\}. \label{eq:WTE_RTA_general}
\end{split}
\end{equation}
Consequently, the local temperature $T_l(\R,t)$ is no longer an input and we solve for it self-consistently to enforce energy conservation via
\begin{equation}
\sum_{{s}{\q}} \frac{\hbar\omega_{s}}{\mathcal{V}N_\mathrm{C}}\left(\Ntcomp_{s,s}(\q)-\delta(\omega)\delta(\k)\bar{\tenscomp{N}}^{T_0}_s({\q})\right) = C\Delta T(\k,\omega) , \label{eq:energy_conservation}
\end{equation}
with the total heat capacity $C = \sum_{s,\q} c_s(\q)$.
More details on the implementation can be found in Appendix~\ref{sec:implementation}.

\section{Numerical Results}
As a benchmark, we start by calculating static ($\omega=0$) thermal conductivities of silicon, \CPB and \LZO using the gradient source term $\PtgradT$ from Eq.~\ref{eq:gradT_source} in the macroscopic regime with a macroscopic grating period, before turning to dynamical thermal conductivity and microscopic grating sizes that simulate, for example, geometries of transient thermal grating experiments.
Note that all thermal conductivities are reported along the thermal grating.
We use the same unit cell convention as~\cite{simoncelli2022wigner} and report $\kappa=\kappa^{xx}$ along the direction of the shortest unit cell axis in \CPB.
The results are shown in Fig.~\ref{fig:reproduction}\,a.
In silicon, we observe that contributions from coherences are negligible compared to populations and the thermal conductivity is in good agreement with literature.
Inter-branch degeneracies are rare and the velocity operator is close to diagonal in the phonon eigenbasis, so the anticommutator drive creates a mostly diagonal response.
Off-diagonal coherences dephase rapidly due to the commutator term $\i [\W, \Nt]$ at splittings $\Delta \omega_{ss'}$ that far exceed the damping rate $\frac{\Gamma_s + \Gamma_{s'}}{2}$.
In \CPB and \LZO, by contrast, dense clusters of flat optical branches create many quasi-degenerate pairs $(s,s')$ with small $\Delta \omega_{ss'}$.
The source's anticommutator directly pumps coherences, and the RTA damping competes less effectively when $\Delta \omega_{ss'} \sim \frac{\Gamma_s + \Gamma_{s'}}{2}$.
Therefore, the coherence contribution is significant and even becomes the dominant contribution to the thermal conductivity above 240 and 800\,K, respectively.
The total thermal conductivity exactly matches the results obtained from \textsc{phono3py}~\cite{caldarelli2022many} and the good experimental agreement of those results has been previously discussed.
In \LZO, the contributions from populations and coherences are also in good agreement with previous results; however, in \CPB we observe a minor deviation, with a larger contribution from coherences and a smaller contribution from populations across the whole temperature range~\cite{simoncelli2019unified,caldarelli2022many}.

\begin{figure*}[t]
    \centering
    \includegraphics{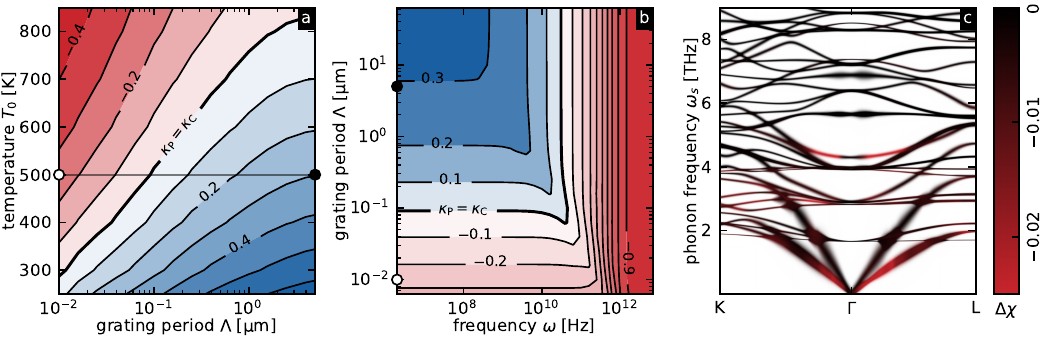}
    \caption{\textbf{a,b}\,Two contour plots that sample the transport character $\chi(T,k,\omega)$ of \LZO at two different planes. The thick black contours indicate equal contributions from populations and coherences ($\chi=0$). \textbf{a} shows the plane for the static case.  \textbf{b}\, shows the plane for a temperature of 500\,K. The intersecting line of the planes is represented by a line that starts with an open circle and ends with a solid circle in a and b. \textbf{c}\, Phonon dispersion of \LZO along the $\overline{\mathrm{K}\Gamma\mathrm{L}}$ high-symmetry path. Encoded in the color of the phonon modes is the change in transport character $\Delta\chi$ between the static and the high-frequency case. The thickness of each phonon branch is proportional to the square root of its model contribution to the total thermal conductivity, i.e. line width $\propto\sqrt{\kappa_{\q s}}$. Note that the energy axis is truncated at 9\,THz, because the high-energy optical phonons show no changes in $\chi$.}
    \label{fig:dynamical_kappa}
\end{figure*}

Further, we can study size effects by varying the grating period $\Lambda=2\pi/k$.
Finite $\k$ filters carriers by mean free path: modes with $\Lambda_\mathrm{mfp} \gg \Lambda$ cannot support the spatial modulation and are suppressed by the convective term $\tfrac{\i\k}{2}\left\{\V,\cdot\right\}$, even at $\omega=0$.
We note that the size effect studied here arises from the size of the thermal drive and is distinct from boundary scattering of finite-sample boundary conditions, which would further reduce the thermal conductivity.

In \LZO, populations and coherences both contribute significantly to the thermal conductivity in the bulk.
The population contribution mainly stems from a few long mean free path acoustic phonon modes close to the Brillouin zone center in the static case.
As the grating period is reduced, these modes are suppressed and the population contribution to the thermal conductivity drops quickly, starting at around 10\,\textmu m.
Interpreting size-induced suppression of coherences requires recognizing that coherences are not associated with a single-mode mean free path. Instead, their contribution is controlled by a mode-pair coherence propagation length, which is set by a combination of finite lifetimes and frequency mismatch between the participating modes.
In practice, our WTE results show that for the materials and conditions considered here, coherence contributions remain largely unchanged as the grating period is reduced, until suppression becomes apparent only for periods around 50\,nm.

The results for room temperature and 700\,K are shown in Fig.~\ref{fig:reproduction}\,b.
Note that the onset of the size effect shifts to smaller grating periods at higher temperatures as phonon mean free paths decrease with increasing temperature.
In \CPB, the phonon mean free paths of the zone-center acoustic modes are only an order of magnitude larger than the mean free paths of higher-frequency modes~\cite{simoncelli2022wigner}.
Size effects start to appear on the scale of 50\,nm for both populations and coherences.
With extreme ultraviolet thermal grating techniques reaching grating periods in the sub-50\,nm regime, these size effects are experimentally accessible~\cite{bencivenga2015fourwave,chergui2023progress}.

Having investigated thermal conductivities and size effects under static conditions, we now move on to dynamical thermal conductivities.
From molecular dynamics simulations~\cite{volz2001thermal} and linearized BTE calculations~\cite{chaput2013direct}, it is understood that the population thermal conductivity is strongly suppressed as the excitation frequency approaches the material's phonon lifetimes.

Figure~\ref{fig:reproduction}\,c shows the dynamical thermal conductivity of \LZO at 500\,K for two different grating periods.
For the bulk grating period of 63\,\textmu m, we observe that the real part of the thermal conductivity starts to decrease at around 1\,GHz, which coincides with the onset of a peak in the imaginary part.
The peak marks a $\pi/2$ phase lag of the heat flux behind the drive and occurs where $\omega$ matches the effective relaxation rate of the dominant heat-carrying phonons.

To quantify the character of the thermal conductivity, we define a dimensionless parameter
\begin{equation}
\chi = \frac{\kappa_\mathrm{P} - \kappa_\mathrm{C}}{\kappa_\mathrm{P} + \kappa_\mathrm{C}},
\end{equation}
which we call the transport character.
Fully population-dominated transport $\chi$ takes a value of 1, fully coherence-dominated transport $-$1 and 0 for equal contributions from populations and coherences.
Figure~\ref{fig:dynamical_kappa}\,a presents a contour plot of $\chi$ as a function of grating period and temperature for \LZO.
The thick black contour indicates equal contributions from populations and coherences.
One can see that coherences dominate the thermal conductivity at experimentally accessible grating periods on the micron scale at moderately elevated temperatures.
A second cut through the $\chi(T,k,\omega)$ space is shown in Fig.~\ref{fig:dynamical_kappa}\,b, where the temperature is fixed at 500\,K.
The intersecting line of the two planes in Figs.~\ref{fig:dynamical_kappa}\,a and b is represented by a line that starts with an open circle at small grating periods and ends with a solid circle at large grating periods in both figures.
As is seen in the BTE, increasing $\k$ (space filter) removes long-mean-free-path population channels; increasing $\omega$ (time filter) removes long-lifetime population channels.
Coherence channels---fed by many near-degenerate optical pairs with finite $\V_{ss'}$---survive both filters longer, so $\chi$ drifts negative and curves collapse onto a coherence-dominated tail, emerging as vertical parallel contour lines in the high-$\omega$ region of Fig.~\ref{fig:dynamical_kappa}\,b.

To further investigate the origin of the change in thermal transport character, we compare two distinct scenarios.
The static case serves as a baseline with a large grating period at 500\,K, where $\kappa_\mathrm{P}=2\kappa_\mathrm{C}$ or $\chi=\frac{1}{3}$.
This baseline is compared against the high-frequency case at 31\,GHz, where $\kappa_\mathrm{P}=\kappa_\mathrm{C}$ or $\chi=0$.
The phonon dispersion of \LZO along the $\overline{\mathrm{K}\Gamma\mathrm{L}}$ high-symmetry path is shown in Fig.~\ref{fig:dynamical_kappa}\,c.
Along with the dispersion, the change in transport character $\Delta\chi$ between the static and the high-frequency case is encoded in the color of the phonon modes.
At no point in the Brillouin zone do we observe a positive change in $\chi$, which would indicate a shift towards population-dominated transport.
Only in the acoustic and some low-lying optical modes do we observe a small shift towards coherence-dominated transport of up to $\Delta\chi=-0.03$.
We, therefore, conclude that the change in thermal conductivity is not due to phonon modes shifting their transport character, but rather to the suppression of population-dominated contributions to the thermal conductivity.
For comparison, the same figure for the absolute $\chi$ of the macroscopic, static baseline is shown in Appendix~\ref{sec:baseline_character}.
We further suppressed the population contribution until reaching $\chi=0$ by decreasing the grating period to 90\,nm.
The general picture remains unchanged, while the changes in $\chi$ are even smaller.

\section{Conclusion}
We developed a frequency-domain formulation of the phonon Wigner transport equation that incorporates general space-time sources and yields a linear problem.
In this framework, a gradient-type drive produces the dynamical thermal conductivity $\kappa(\k,\omega)$ without requiring a self-consistent temperature field, while arbitrary sources are accommodated through an energy-conserving closure.
We implemented two numerical strategies to solve the resulting linear problem, based on either direct inversion or iterative solvers.
Applied to \CPB and \LZO, our framework predicts static and dynamical thermal conductivities in addition to size effects.
Based on our calculations, we predict deviations from bulk diffusivities at experimentally accessible length and time scales.
The decomposition of the thermal conductivity into populations and coherences across the Brillouin zone reveals that the character of thermal transport within a phonon branch (i.e., population dominated or coherence dominated) does not depend on the grating period or drive frequency.

We note that the smallest grating periods explored here, spanning approximately 50 unit cells, push towards the limits of the assumption of temperature gradients larger than the unit cell---a similar caveat applies to BTE-bases nonlocal approaches.
Including them is still of value as transient grating techniques are moving into this regime:
EUV\cite{Foglia2023Extreme} and hard x-ray\cite{Li2026Nanoscale} implementations now reach nanoscale periods, with reported limits down to 10\,nm.

Beyond this work, explicit Green's function inversion enables systematic studies of system responses to a variety of source terms and to different forms of the collision operator.
For instance, it would be of interest to consider a collision operator that is not restricted to the annihilation of coherences. Additionally, specific parts of the Brillouin zone can be targeted, where the results of electron-phonon coupling studies can be used to construct source terms that mimic the effect of ultrafast laser pulses.

\section{Acknowledgements}
L.K. acknowledges support from a Fonds de Recherche du Québec---Nature et Technologies (FRQNT) Merit fellowship. B.J.S. acknowledges support from the NRC Quantum Sensors Challenge Program and the Canada Research Chairs program. S.H. acknowledges support from the NSERC Discovery Grants Program under Grant No. RGPIN-2021-02957 and FRQNT Nouveau Chercheur No. 341503.

\section{Data Availability}
The data that support the findings of this article are openly available. The code used to produce the results in this work is available at~\cite{greenWTE,phono3py} and input data are available at~\cite{ifc_reference}.

\bibliography{main}

\appendix
\section{Vectorization operator}\label{sec:vectorization}
The vectorization operator $\vecop(A)$ stacks the columns of an $(m\times n)$ matrix $A$ on top of one another, resulting in an $(mn\times1)$ vector:
\begin{equation}
    \vecop(A) = \left[ a_{1,1}, \dots, a_{m,1}, \dots, a_{m,2}, \dots, a_{1,n}, \dots, a_{m,n}\right]^\mathrm{T} \label{eq:vecop}
\end{equation}
With the Kronecker product $\otimes$ and the identity matrix $\I$, we can vectorize as
\begin{align}
    \vecop(ABC) &= (C^\mathrm{T}\otimes A) \vecop(B). \label{eq:vecop_application1}\\
    \vecop(AB) &= (\I\otimes A) \vecop(B) = (B^\mathrm{T} \otimes \I) \vecop(A). \label{eq:vecop_application2}
\end{align}
The transpose $A^\mathrm{T}$ of a matrix $A$ is denoted with a superscript T.

\begin{figure*}[t]
    \centering
    \includegraphics{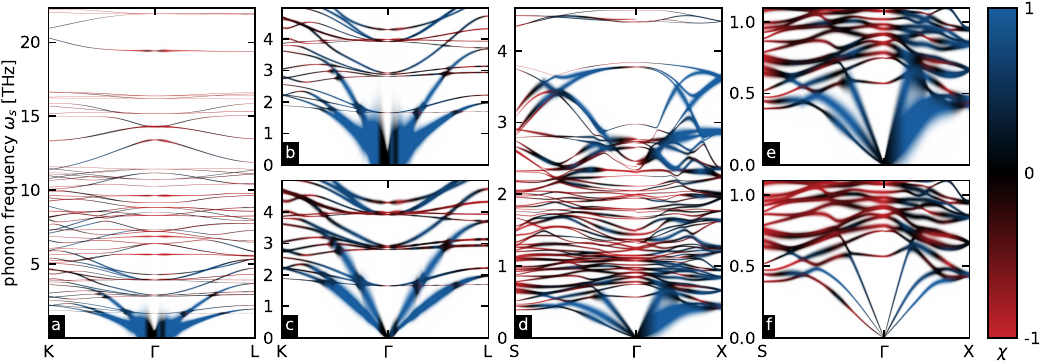}
    \caption{Phonon dispersion of \LZO (\CPB) along the $\overline{\mathrm{K}\Gamma\mathrm{L}}$ ($\overline{\mathrm{S}\Gamma\mathrm{X}}$) high-symmetry path. Encoded in the color of the phonon modes is the transport character $\chi$ in the static case for a macroscopic grating period \textbf{a}, \textbf{b}, \textbf{d}, and \textbf{e} and at $\omega$=31 (87)\,GHz \textbf{c}, \textbf{f}. The thickness of each phonon branch is proportional to the square root of its model contribution to the total thermal conductivity, i.e., line width $\propto\sqrt{\kappa_{\q s}}$. Panel \textbf{a,d} shows the full energy range, while panel \textbf{b}, \textbf{c}, \textbf{e}, and \textbf{f} is a zoom into the low-energy part of the spectrum.}
    \label{fig:baseline_character}
\end{figure*}

\section{Implementation}\label{sec:implementation}
The implementation for solving the WTE for a distribution $\Nt$ for a given $\Delta T$ remains unchanged for all sources.
We make use of the independence between different points in the Brillouin zone in the RTA to construct $N_\mathrm{q}$ independent linear systems of size $(N_\mathrm{ph}\times N_\mathrm{ph})$.
$N_\mathrm{q}$ is the number of $\q$-points in the Brillouin zone and $N_\mathrm{ph}$ is the number of phonon branches.
Making use of the vectorization operator defined in Eqs.~\ref{eq:vecop}--\ref{eq:vecop_application2}, we can rewrite Eq.~\ref{eq:WTE_RTA_general} as $N_\mathrm{q}$ independent linear systems of equations:
\begin{widetext}
\begin{equation}
\left[ \i \left( \left( \I \otimes \W \right) - \left( \W \otimes \I \right) - \omega \I \right) - \frac{\i k}{2} \left( \left( \I \otimes \V^\beta \right) + \left( \left(\V^\beta \right)^\mathrm{T} \otimes \I \right) \right) + \frac{1}{2} \left( \left( \I \otimes \G \right) + \left( \G \otimes \I \right) \right) \right] \vecop(\Nt) = \vecop(\Pt + \G \Nteq{T_l}). \label{eq:WTE_RTA_vec}
\end{equation}
\end{widetext}
$\I$ is the identity matrix, $\otimes$ is the Kronecker product, and the resulting linear system has a size of $(N_\mathrm{ph}^2\times N_\mathrm{ph}^2)$.
From this equation, two approaches can be taken to obtain $\Nt$ and $\Delta T$.

Two solver strategies are implemented in the code.
The Green's function can be computed explicitly for a $(\k,\omega)$ pair by inverting the left-hand-side operator of Eq.~\ref{eq:WTE_RTA_vec}.
An advantage of this approach is that the Green's function only depends on the material properties and not on the source term $\Pt$.
The computational cost of a full inversion of $\Lint$ scales with $\mathcal{O}(N_\mathrm{q} N_\mathrm{ph}^6)$.
Complex unit cells with many phonon branches can make this approach infeasible, but dense sampling of the Brillouin zone is still possible.
Applying the Green's function to a source term is then a simple matrix-vector multiplication that is computationally inexpensive.
That makes the explicit Green's function approach very efficient if the response to a variety of source terms is of interest, the material studied has a simple unit cell or many iterations are needed to converge $\Delta T$.

The second option is to use a direct solver on the full system of equations.
Direct inversion scales much better than full inversion.
However, the full system depends on the source term and must be reconstructed for every new source.
In addition to that, should an iterative approach be needed to converge $\Delta T$, the full system must be re-constructed and solved in every iteration.
The code supports GPU implementations of the GMRES solver with a Jacobi preconditioner and dense LU decomposition, both of which are implemented in the \textsc{CUDA Toolkit}.
Solutions from nearby $\omega$ and previous iterations over $\Delta T$ and $\Nt$ are used to accelerate convergence.
Direct inversion is a good choice for complex unit cells with many phonon branches and when only a few source terms are of interest.

When using a source term that requires self-consistent calculation of $\Delta T$, closure is imposed by Eq.\ref{eq:energy_conservation}.
Depending on system stability and required convergence speed, the code allows one to choose between a simple fixed-point iteration, an Aitken-accelerated fixed-point iteration, or MINPACK's HYBRD method on the complex temperature plane.
In this context, a calculation is considered converged when $\Delta T$ computed from the closure Eq.~\ref{eq:energy_conservation} in two subsequent iterations $k-1, k$ fulfills
\begin{equation}
|\Delta T_{k-1} - \Delta T_k| < \tau_\mathrm{abs} + \tau_\mathrm{rel} \cdot \max(|\Delta T_{k-1}|, |\Delta T_k|, 1), \label{eq:convergence_criterion}
\end{equation}
where $\tau_\mathrm{abs}$ and $\tau_\mathrm{rel}$ are absolute and relative tolerances that can be set by the user.
The relative tolerance is also passed to the GMRES solver if that is used.
As a diagnostic, the relative change in the Euclidean norm of the Wigner distributions is tracked between iterations via $\eta_k = \frac{||\Nt_k - \Nt_{k-1}||_2}{||\Nt_{k-1}||_2}.$

Going beyond the RTA to a full scattering matrix approach to describe the phonon-phonon interactions is technically possible, but Eq.~\ref{eq:WTE_RTA_vec} will no longer be a system of $N_\mathrm{q}$ independent linear systems of equations, because the scattering matrix will allow interdependence between the systems.
One will end up with a very large coupled system that will require careful tuning in order to be efficiently solved.
Considering only a subspace across the most important phonon modes and materials with fewer atoms in the unit cell are promising strategies to implement this approach beyond the RTA.

\section{Computational Details}
The code for performing the calculations presented here is published as a Python package under the name \textsc{greenWTE}.
The source code and documentation with examples can be found at~\cite{greenWTE}.
All threshold parameters for the iterative solver were set to $\tau_\mathrm{rel} \le 10^{-12}$ and $\tau_\mathrm{abs} = 0$.
Second- and third-order interatomic forces for silicon were taken from the \textsc{phono3py} examples~\cite{phono3py}.
We subsequently used \textsc{phono3py} to compute the phonon properties required as input to \textsc{greenWTE} on a $15\times15\times15$ grid.
In the case of \CPB and \LZO, precomputed interatomic force constants from~\cite{ifc_reference} were used to compute the phonon properties on $8\times8\times5$  and $19\times19\times19$ grids, respectively.
All calculations use the smooth phase convention for the computation of the dynamical matrices and derived quantities to ensure invariance to the choice of unit cell and size consistency.

\section{Mode-resolved transport character}\label{sec:baseline_character}
For the baseline reference for Fig.~\ref{fig:dynamical_kappa}\,c, the thermal conductivity of \LZO was calculated at a background temperature of 500\,K, with a grating period of 63\,\textmu m in the static case, $\omega=0$.
The mode-resolved transport character $\chi$ is plotted across the full energy range across the $\overline{\mathrm{K}\Gamma\mathrm{L}}$ path in Fig.~\ref{fig:baseline_character}\,a.
From the thickness of the lines it is apparent that the population contribution to the thermal conductivity stems from a few low-energy zone-center acoustic modes, while the coherence contribution is more evenly distributed across the optical phonon modes.
The total transport character is $\chi=\frac{1}{3}$.
A zoom into the low-energy part of the spectrum is shown in Figs.~\ref{fig:baseline_character}\,b and c.
One can directly compare the static case with the high-frequency case at 31\,GHz that was discussed in the main text.
While the character of the individual phonon modes does not change significantly, the population-dominated acoustic modes are suppressed in the high-frequency case and the total transport character shifts to $\chi=0$.

In addition to \LZO, Figs.~\ref{fig:baseline_character}\,d--f show the corresponding phonon dispersion and mode-resolved transport character of \CPB along the $\overline{\mathrm{S}\Gamma\mathrm{X}}$ path for a background temperature of 100\,K.
Qualitatively, we observe the same suppression of thermal conductivity in the low-frequency modes with strong population character and overall stable transport characters across the Brillouin zone.  
\end{document}